# Neutron scattering studies of complex lattice dynamics in energy materials[*]

REN Qingyong[1, 2, 3, *], WANG Jianli[4], LI Bing[5], MA Jie[6], TONG Xin[1, 2, 3]

1. Spallation Neutron Source Science Center, Dongguan 523803, China
2. Institute of High Energy Physics, Chinese Academy of Sciences, Beijing 100049, China
3. Guangdong Provincial Key Laboratory of Extreme Conditions, Dongguan 523803, China
4. Center for Neutron Scattering and Advanced Light Sources, Dongguan University of Technology, Dongguan 523000, China
5. Institute of Metal Research, Chinese Academy of Sciences, Shenyang 110016, China
6. School of Physics and Astronomy, Shanghai Jiao Tong University, Shanghai 200240, China

**Abstract**

Lattice dynamics play a crucial role in understanding the physical mechanisms of cutting-edge energy materials. Many excellent energy materials have complex multiple-sublattice structures, with intricate lattice dynamics, and the underlying mechanisms are difficult to understand. Neutron scattering technologies, which are known for their high energy and momentum resolution, are powerful tools for simultaneously characterizing material structure and complex lattice dynamics. In recent years, neutron scattering techniques have made significant contributions to the study of energy materials, shedding light on their physical mechanisms. Starting from the basic properties of neutrons and double differential scattering cross sections, this review paper provides a detailed introduction to the working principles, spectrometer structures, and functions of several neutron scattering techniques commonly used in energy materials research, including neutron diffraction and neutron total scattering, which characterize material structures, and quasi-elastic neutron scattering and inelastic neutron scattering, which characterize lattice dynamics. Then, this review paper presents significant research progress in the field of energy materials utilizing neutron scattering as a primary characterization method. 1) In the case of $Ag_8SnSe_6$ superionic thermoelectric materials, single crystal inelastic neutron scattering experiments have revealed that the "liquid-like phonon model" is not the primary contributor to ultra-low

---

lattice thermal conductivity. Instead, extreme phonon anharmonic scattering is identified as a key factor based on the special temperature dependence of phonon linewidth. 2) Analysis of quasi-elastic and inelastic neutron scattering spectra reveals the changes in the correlation between framework and $Ag^+$ sublattices during the superionic phase transition of $Ag_8SnSe_6$ compounds. Further investigations using neutron diffraction and molecular dynamics simulations reveal a new mechanism of superionic phase transition and ion diffusion, primarily governed by weakly bonded Se atoms. 3) Research on $NH_4I$ compounds demonstrates a strong coupling between molecular orientation rotation and lattice vibration, and the strengthening of phonon anharmonicity with temperature rising can decouple this interaction and induce plastic phase transition. This phenomenon results in a significant configuration entropy change, showing its potential applications in barocaloric refrigeration. 4) In the $CsPbBr_3$ perovskite photovoltaic materials, inelastic neutron scattering uncovers low-energy phonon damping of the $[PbBr_6]$ sublattice, influencing electron-phonon coupling and the band edge electronic state. This special anharmonic vibration of the $[PbBr_6]$ sublattice prolongs the lifetime of hot carriers, affecting the material's electronic properties. 5) In MnCoGe magnetic refrigeration materials, *in-situ* neutron diffraction experiments highlight the role of valence electron transfer between sublattices in changing crystal structural stability and magnetic interactions. This process triggers a transformation from a ferromagnetic to an incommensurate spiral antiferromagnetic structure, expanding our understanding of magnetic phase transition regulation. These examples underscore the interdependence between lattice dynamics and other degrees of freedom in energy conversion and storage materials, such as sublattices, charge, and spin. Through these typical examples, this review paper can provide a reference for further exploring and understanding the energy materials and lattice dynamics.

# 1. Introduction

The microscopic crystal structure of solid matter serves as the foundation for understanding macroscopic physical phenomena. However, crystal many alone fail to fully explain many such behaviors. According to quantum mechanics and solid-state theory, the constituent units (atoms, ions, or molecules) in solids vibrate within a potential well generated by their neighboring units. These vibrations exhibit quantized energy signatures, resulting in quasiparticles called phonon. The collective vibrational patterns between these constituent units give rise to a specific relationship between phonon energy and momentum, known as the phonon dispersion or phonon spectrum[1]. Phonons encode both kinetic and potential energies of the lattice, making them critical for interpreting thermal properties, including internal energy, specific heat, phase transition, thermal conductivity, thermal expansion, etc.[2] These properties underpin the functionality of many cutting-edge energy materials, such as phase-transition refrigeration and energy storage, thermoelectric conversion, battery technologies, and photovoltaics (see Fig. 1)[3-6]. Consequently, lattice dynamics – the study of atomic vibration behavior - has emerged as a pivotal research direction for optimizing or designing energy materials.

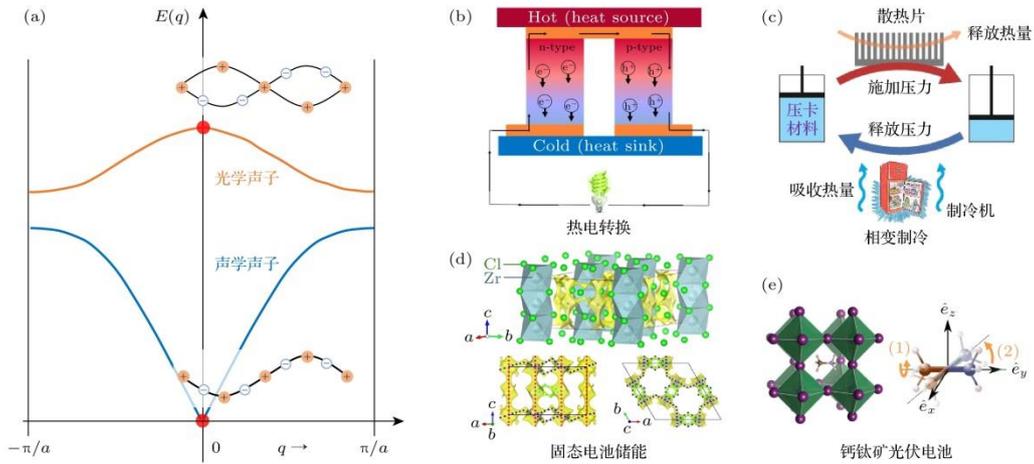

**Figure 1.** Lattice dynamics and energy materials: (a) Lattice vibrations and phonon dispersion; (b) thermoelectric materials and transport properties[5]; (c) barocaloric effect and phase transition entropy change[6]; (d) solid-state electrolyte and ionic diffusion[4]; (e) solar cell and lattice vibrations in perovskites[3].

Taking semiconductor thermoelectric materials as an example, good performance requires both good electrical transport properties and as low as possible lattice thermal conductivity[7]. In electrical transport, many scattering potentials affecting carrier mobility are closely related to lattice vibration, including acoustic phonon deformation potential, optical phonon deformation potential, piezoelectric scattering, and polar optical phonon scattering[8]. For lattice thermal conductivity, phonons serve as primary heat carriers, and their group velocity, lifetime (or mean free path), and density of states collectively determine thermal transport efficiency[9–12]. These dual considerations underscore the critical role of lattice dynamics in advancing thermoelectric materials.

Recent research on energy materials has gradually shifted toward complex crystal structure systems, and the lattice dynamics of these materials has also become extremely complex[12–16]. Most thermoelectric materials with intrinsically low thermal conductivity often feature structural complexities such as ultra-large unit cells[17–19], a large number of ordered or disordered vacancy defects[20,21], local distortion or disorder[22–24], host-guest atoms or molecules[25,26], ferroelectric or chemical instability, multiple sublattices, etc.[27,28] Such structural complexity and diversity make the phonon spectra of related materials also extremely complex, such as ultra-low acoustic phonon cut-off frequency[18], strong anharmonicity[29,30], obvious phonon softening or overdamping[31,32], the "waterfall phenomenon" of optical phonon softening at the Brillouin zone center[28,33], or an "avoided crossing" between acoustic and optical branches[25,34]. In superionic solid electrolyte, ions of some sublattices exhibit long range diffusion behaviors, and these diffusion behaviors are closely related to the thermal vibrations of other sublattices[35–39]. In photovoltaic perovskite materials, polyhedrons in the lattice exhibit giant anharmonic vibrations, while organic molecules exhibit local rotational motions. These complex lattice dynamics are the important physical basis for driving the structural phase transition of materials, and also endow related materials with efficient hot carrier transport properties[3,40,41]. In phase-transition refrigeration or energy storage materials, lattice entropy change is an important contribution to the total entropy change, and the local rotation of molecules endows lattice entropy with a very large configurational entropy[42].

Complex crystal structure and lattice dynamics constitute an important physical basis for many new energy materials, but at the same time, it also brings great challenges to the understanding of the physical mechanism of related materials. As the first elementary excited quasiparticle introduced into physics, lattice vibration phonon has been studied for more than 90 years[43]. It has developed a wealth of research methods, including theoretical simulation and experimental characterization. Especially in theoretical calculation, from density functional theory to machine learning molecular dynamics, lattice dynamics simulation can accurately reproduce experimental results[12]. However, with the complexity of the crystal structure and lattice dynamics of novel energy materials, understanding lattice dynamics through theoretical simulation is facing great challenges. These challenges come not only from the construction of theoretical models, but also from the lack of accurate grasp of the relevant microscopic physical pictures. Under this circumstance, it is very important to directly characterize lattice dynamics by experimental means, including Raman, Brillouin scattering and neutron[44,45].

This article reviews and discusses the neutron scattering technologies and some important achievements in the study of crystal structures and lattice dynamics of energy materials. Firstly, the key technologies and advantages of neutrons in the characterizations of crystal structures and lattice dynamics of functional materials are introduced; then, taking superionic thermoelectric materials as an example, the success of neutron scattering technique in the study of ultra-low lattice thermal conductivity mechanism is introduced; taking solid electrolytes as an example, the key interactions between ion diffusion and lattice vibration is revealed; taking plastic crystal materials as an example, the intrinsic relationship between plastic crystal phase transition, configuration entropy and barocaloric refrigeration is revealed; finally, the interaction between lattice vibration, charge and spin is introduced to provide a basis for understanding more energy materials.

## 2. Neutron characterization method for lattice dynamics

2.1 Basic characteristics of neutron

The importance of neutron scattering technology in the study of material structure and lattice dynamics is mainly due to the special properties of neutrons[46,47]. 1) Neutrons are electrically neutral particles (Fig. 2 (a)) with a large penetration depth. They can be used to characterize the internal or bulk information of materials or to conduct non-destructive testing. At the same time, they are easy to load various extreme sample environments, including ultra-low temperature cryostats, extremely high temperature furnaces, strong magnets, pressure cells, etc., to meet the needs of various scientific researches. 2) Neutrons are mainly scattered by atomic nucleus, unlike X-ray or electron beams, which mainly interact with the electron cloud outside the nucleus (Fig. 2(b)). The different ways of interaction modes with matter determine that the scattering cross section of X-ray or electron beams of each element is proportional to the atomic number, while the scattering cross section of neutron is independent of the atomic number and varies greatly between adjacent elements (Fig. 2 (c))[47,48]. Neutrons can measure light elements, such as H and Li, which are important elements in energy materials. At the same time, neutrons can distinguish between elements and isotopes with adjacent atomic numbers, which is complementary to X-rays. 3) Neutron is composed of one up quark and two down quarks. This structure endows neutrons with a spin moment, enabling it magnetically scatter with unpaired electrons outside the nucleus. This has unique advantages in magnetic structure analysis and spin excitation measurement. 4) The mass of neutrons is very large (about 1842 electron masses), and neutrons with wavelengths at the atomic scale (in the order of angstroms) have much lower energy than the corresponding X-rays, comparable to the elementary excitation energy in solid matter. Neutron scattering (compared to light scattering) can cover a large range of energy and momentum, and has relatively high energy and momentum resolution, making it one of the best methods for simultaneously characterizing material structure and measuring lattice dynamics. It can be used to study various lattice dynamics behaviors such as ion diffusion, molecular rotation, lattice collective vibration (phonons), crystal fields, etc. (Fig. 2(d))[49], providing a more intuitive physical picture for the understanding of the physical properties of materials[44,45,50]. This low energy feature also makes the perturbation of neutrons on solid matter close to the equilibrium state, which conforms to the linear response approximation, but this approximation does not always hold in X-ray scattering experiments.

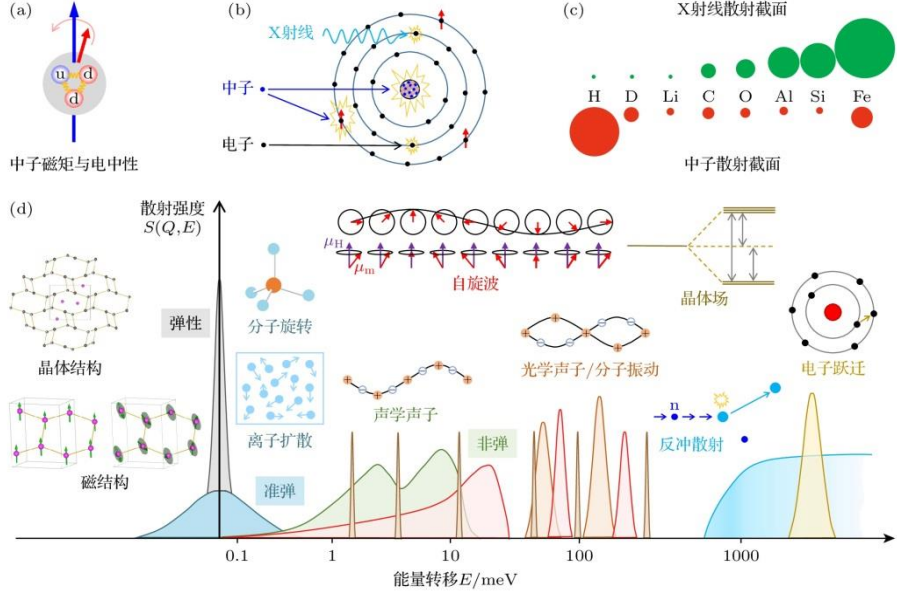

**Figure 2.** Overview of neutron scattering technology: (a) Neutron is a neutral particle with a magnetic moment; (b) the interactions of neutrons, electrons and X-rays with matter; (c) neutron and X-ray cross sections for some typical elements[47,48]; (d) the energy range spanned by the elastic, quasi-elastic and inelastic neutron scattering technologies and the typical physical contents in the energy range[49].

According to Fermi's golden rule, the information directly obtained by neutron scattering experiments is the double differential scattering cross section:

$$\frac{\mathrm{d}^2\sigma}{\mathrm{d}\Omega \mathrm{d}E} = \frac{\sigma}{4\pi}\frac{\boldsymbol{k}_\mathrm{f}}{\boldsymbol{k}_\mathrm{i}}NS(\boldsymbol{Q},\omega), \tag{1}$$

Among, $\sigma$ is the neutron scattering cross section of the element in the sample, $\boldsymbol{k}_\mathrm{i}$ and $\boldsymbol{k}_\mathrm{f}$ are the incident and scattered neutron wave vectors, $N$ is the number of atoms. $S(\boldsymbol{Q},\omega)$ is the dynamical structure factor, inward from the underlying statistical theory aspect can be expressed as:

$$S(\boldsymbol{Q},\omega) = \frac{1}{2\pi\hbar}\iint G(\boldsymbol{r},t)\mathrm{e}^{\mathrm{i}(\boldsymbol{Q}\cdot\boldsymbol{r}-\omega t)}\mathrm{d}\boldsymbol{r}\mathrm{d}t, \tag{2}$$

$$G(r,t) = \left\langle \sum_{i=1}^{N}\sum_{j=1}^{N} \delta[r' + r - r_i(t)]\delta[r - r_j(0)] \right\rangle, \tag{3}$$

where, $G(r,t)$ is the van Hove correlation function, which describes the correlation between atom $j$ at position $r$ at time $t = 0$ and atom $i$ at position $r' + r$ at time $t$. $G(r,t)$ involves the correlation between time and space. Outward to the level of condensed matter physics, $S(Q,\omega)$ can be described as:

$$\begin{aligned} S(Q,\omega) = \quad & \frac{1}{2NM} e^{-Q^2\langle u^2\rangle} \sum_{jq} \frac{|Q \cdot \varepsilon_j(q)|^2}{\omega_j(q)} \\ & \times \left[n(\omega) + \frac{1}{2} \pm \frac{1}{2}\right] \times \delta(Q \pm q - \tau) \\ & \times \delta[E_f - E_i \pm \hbar\omega(q)] \end{aligned} \tag{4}$$

This equation represents the process by which a neutron transfers energy to a sample and creates (+) or annihilates (−) a phonon in the sample. $Q = k_i - k_f$ is the scattering vector, $\tau$ is the reciprocal vector of the lattice, $q$ and $\hbar\omega(q)$ are the wave vector and energy of the excited quasiparticle, two Dirac ($\delta$) functions correspond to the Fourier transformation of position $r$ and time $t$ in Eq. (3), and constrain the scattering event to obey the conservation of momentum and energy[46,47,51]. $\varepsilon_j(q)$ is the polarization vector of the phonon normal mode, which can be used to distinguish the transverse phonon from the longitudinal phonon in the experimental measurement. $e^{-Q^2\langle u^2\rangle}$ is the Debye-Waller factor.

According to whether energy transfer occurs during the scattering of neutron with matter, that is, whether $\hbar\omega(q)$ or $E_f - E_i$ in Eq. (4) is zero, neutron scattering can be classified into elastic scattering and inelastic scattering (Fig. 2(d)), corresponding to different types of spectrometers and measurement information. Elastic scattering includes neutron diffraction, total scattering, diffuse scattering, small angle neutron scattering, neutron reflection, etc., while inelastic scattering includes quasi-elastic neutron scattering and inelastic neutron scattering.

2.2 Elastic scattering and micro crystal structure characterization.

1) Neutron diffraction spectrometer and principle. Neutron diffraction is the most commonly used method in elastic scattering technology. Similar to X-ray diffraction and electron diffraction, neutron diffraction is also a common means to characterize the crystal structure of materials, which follows Bragg's law of diffraction. At present, there are mainly two types of neutron sources, one is continuous neutron source (mostly reactor neutron sources, such as ILL in France, NIST and HFIR in the United States, FRM II in Germany, ANSTO in Australia, and CMRR and CARR in China), and the other type is pulsed neutron source (mostly spallation neutron sources, such as J-PARC in Japan, SNS in the United States, ISIS in the United Kingdom, and CSNS in China). The diffraction spectrometers with the two different neutron sources have different structures and working modes. The diffraction spectrometer with reactor neutron source is similar to a traditional X-ray diffraction spectrometer. It uses a single wavelength neutron and measure different diffraction peaks $d_{hkl}$ by scanning a wide scattering angle $2\theta$ (Fig. 3(a)). Therefore, in this type of fixed-wavelength diffraction experiment, $2\theta$ and $d_{hkl}$ are the two main variables in the Bragg's formula (Fig. 3(b)). The diffractometer with a pulsed neutron source has a different structure. The detector angle of this type of diffractometer is fixed, and the neutron flight distance between the moderator (a device that reduces the high-energy neutrons produced by high-energy proton targeting from MeV level to meV level) and the detector is also fixed (Fig. 3(c)). By measuring the neutron flight time (time of flight, $tof$), the neutron velocity or wavelength or energy can be obtained (Fig. 3(d)). Using a certain bandwidth of neutrons in each pulse, different diffraction peaks can be measured one by one. Therefore, the two variables in the measurement process of a pulsed neutron diffractometer are $tof$ and $d_{hkl}$ (Fig. 3(d)).

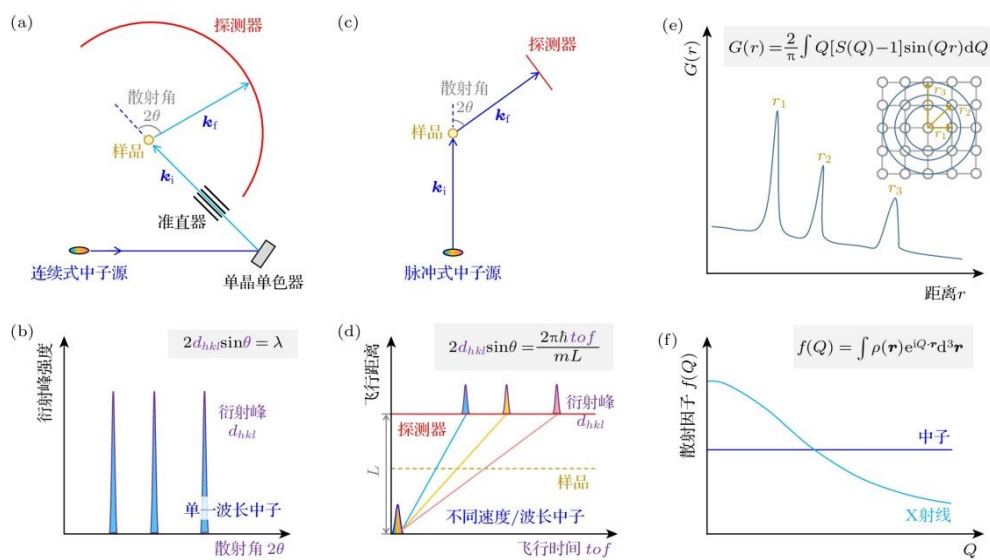

**Figure 3.** Neutron diffraction spectrometers: (a), (b) Structure and working principle of a neutron diffractometer based on a continuous neutron source. (c), (d) Structure of a neutron diffractometer based on a pulsed neutron source and its time-distance diagram or measurement principle. The formats of the Bragg function for these two types of diffractometers are different and their variables are marked with purple color. The blue line in the figure represents the flight path of broadband neutrons, and other colors represent the flight paths of monochromatic neutrons. (e) The neutron total scattering and pair distribution function principle, through the Fourier transform of the structure factor $S(Q)$, the real space distribution, $G(r)$, could be measured. (f) Comparison of atomic scattering factor, $f(Q)$, between X-ray and neutron diffraction. $f(Q)$ for X-ray decays rapidly with momentum transfer $Q$, while $f(Q)$ for neutron remains almost unchanged, so neutron diffraction is more easily to obtain Bragg diffraction peaks at large $Q$. The momentum transfer $Q$ can be obtained by subtract scattered neutron wavevector $k_f$ from the incident neutron wavevector $k_i$, $Q = k_i - k_f$.

2) Advantages of neutron diffraction technology. In the study of energy materials, doping, defects, disordered occupancy, short range order, large unit cell and low symmetry all impose difficulties in structural characterization. Unlike electrons and X-rays, neutrons can detect light elements, distinguish adjacent elements and isotopes, and have obvious advantages in doping and defect analysis. On the other hand, large unit cell and low symmetry mean more and denser Bragg diffraction peaks, and accurate characterization requires high resolution. The low flux and difficulty in focusing determine that the resolution of neutron diffractometers cannot be comparable to X-ray diffractometers, especially synchrotron radiation X-ray diffraction spectrometers, such as the I11 diffractometer at the Diamond Light Source in the UK, which has an optimal resolution of $\Delta d/d \sim 10^{-5}$. However, in recent years, the development of spallation neutron source and decoupled cold-neutron moderator has made high-resolution neutron diffractometers possible, such as the SuperHRPD diffractometer at J-PARC in Japan, which can reach a resolution of $\Delta d/d \sim 10^{-4}$, which can meet the measurement needs of most materials with large unit cells and complex structures.

Another important parameter in crystal structure analysis is the atomic displacement parameters (ADP) or mean-square displacement (MSD), which corresponds to $\langle u^2 \rangle$ in Eq. (4). In diffraction pattern, ADP manifested as the effect of atomic vibration on the scattering factor, that is, the Bragg peak intensity systematically weakens with the increase of $Q$. High quality data acquisition over large $Q$ range can yield accurate three-dimensional ADP, providing atomic-level real-space information for the understanding of lattice dynamics spectroscopy[21]. As mentioned earlier, pulsed neutron diffractometers have relatively higher resolution and can easily cover a larger $Q$ range, making them one of the best means to analyze ADP.

3) Characterization of short-range order and disordered structures. Elastic neutron scattering also provides many solutions for the analysis of disordered structure and short-range order structure. The most widely used one is the neutron total scattering spectrometer. This type of spectrometer uses the pair distribution function (PDF) technique to study the atomic order in real space by Fourier transforming the structure factor $S(Q)$ from the reciprocal space (Fig. 3(e)), providing information from different perspectives for local disorder, lattice distortion, unconventional vibration, etc.[22,52,53] The total scattering neutron spectrometer is essentially a pulsed neutron diffractometer as it can obtain larger $Q$ ($Q = 4\pi \sin\theta/\lambda$) space through ultra-wide neutron bandwidth, and the atomic form factor of neutrons does not decay with the increase of $Q$ (Fig. 3(f)). Therefore, neutrons have more advantages than X-rays in Fourier transform and PDF analysis of $S(Q)$. In the analysis of ion distribution and diffusion channels in solid-state electrolyte materials, an effective method is neutron diffraction combined with maximum entropy method (MEM). Here, the basic principle of information entropy maximization is applied to overcome the limitations of traditional crystallography and effectively resolve the disordered arrangement of ions over a large space[31,39,54]. Neutron diffuse scattering spectrometer also provides another solution for the characterization of short range order or disorder[55]. This technology uses a single crystal sample to observe the diffuse signals caused by short-range order or disorder in the entire three-dimensional reciprocal space, and then uses the inverse Monte Carlo method to restore the arrangement of atoms in real space[56]. In addition, the development of single crystal diffuse scattering technology has also derived 3D PDF technology, which may be able to provide a more comprehensive and direct reconstruction of disorder crystal structure[57].

In addition to neutron diffraction, total scattering and diffuse scattering, elastic neutron scattering technologies also include small angle neutron scattering, neutron reflection, neutron imaging and other techniques. Small angle neutron scattering can be used to characterize nanometer or larger scale inhomogeneous structure and size distribution, such as holes, defects, particles, inner surface[58,59]; Neutron reflectometry can be used to analyze thin film thickness, interface roughness, composition distribution of polymer film etc.[60,61]; and neutron imaging can analyze submicroscopic device structure [62–64].

2.3 Inelastic/quasi-elastic scattering and lattice dynamics.

1) Inelastic/quasi-elastic neutron scattering spectrometer and principle. Inelastic (INS) and quasi-elastic neutron scattering (QENS) processes involve both momentum and energy transfer and are governed by the laws of conservation of momentum and energy. The momentum transfer $Q$ depends on the neutron wave vectors $k_i$ and $k_f$ before and after scattering and the scattering angle $2\theta$ between them, while the energy transfer depends on the energy difference $E_i$-$E_f$ before and after scattering. The scattering angle $2\theta$ is determined by the geometry of the spectrometer and is an easy parameter to determine. However, neutron detectors cannot directly measure the energy of neutrons and need other methods.

There are two main methods for the determination of neutron energy. The first method is to use Bragg's diffraction law to select and determine the wavelength or energy of neutrons by adjusting the diffraction angle of a monochromatic single crystal. This method is usually used for continuous neutron sources. In addition to requiring the rotation axis of the sample to determine the scattering angle $2\theta$, this type of spectrometer also requires two additional rotation axes in front of and behind the sample to operate the rotation angle of the monochromator single crystal and the analyzer single crystal respectively to determine $E_i$ and $E_f$ respectively. Therefore, this type of spectrometer is called a triple-axis spectrometer, see Fig. 4(a).

Another method is to use the neutron time-of-flight (*tof*) through a fixed distance *L* to determine the velocity or energy of the neutron. This method is usually used for pulsed neutron source. In order to facilitate the energy subtraction, one of the parameters in $E_i$ and $E_f$ is required to be a single value. A monochromatic chopper can be placed in front of the sample to make the $E_i$ a single value, or a monochromatic analyzer array can be placed behind the sample to make $E_f$ measured by the detector a single value. The former is called direct geometry inelastic spectrometer (see Fig. 4(c),(d)), and the latter is called indirect/inverted geometry inelastic spectrometer (see Fig. 4(e),(f)).

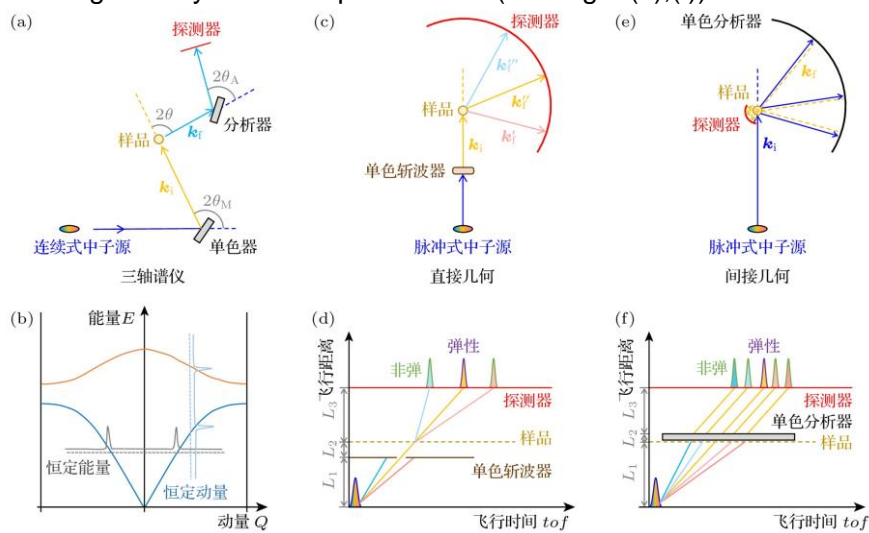

**Figure 4.** Inelastic neutron scattering spectrometer: (a) Schematic diagram of a triple-axis spectrometer based on a continuous neutron source, which has three independently rotating axis, corresponding to the monochromator, the sample stage, and the analyzer. (b) Two methods for a triple-axis spectrometer to scan a dispersion, constant-energy scan and constant-momentum scan. (c) Schematic diagram of a direct-geometry inelastic neutron scattering time-of-flight spectrometer based on a pulsed neutron source. (d) The time-distance diagram or measurement principle of a direct geometry inelastic spectrometer, a bunch of single-energy neutrons are selected by a monochromatic chopper from the pulsed white beam. Then the single-energy neutrons will be inelastically scattered by the sample and their energy and speed will become larger or smaller; the neutron energy after scattering can be determined by the neutron flight time through a fixed distance between the sample and the detector; combined with the angle of the scattered neutron and the neutron energy before scattering, the energy transfer and momentum transfer during the inelastic scattering process can be determined. (e) Schematic diagram of an indirect geometry inelastic neutron scattering time-of-flight spectrometer based on a pulsed neutron source. (f) The "time-distance flight diagram" and working principle of the indirect geometry inelastic spectrometer, this type of spectrometer does not have a monochromator before the sample, but a monochromator analyzer is placed after the sample; the blue line in the figure represents the flight path of broadband neutrons, and other colors represent the flight paths of monochromatic neutrons.

In the characterization of phonon dispersion spectrum, since the triple-axis spectrometer has monochromatic energies before and after neutron scattering with sample, it needs to scan along a certain line in the four-dimensional dispersion spectrum (three momentum dimensions and one energy dimension). Generally, there are two ways: constant momentum scanning (constant-$Q$ scanning) and constant energy scanning (constant-$E$ scanning). On the other hand, the pulsed direct/indirect geometry spectrometers only need to monochromatize the energy at one end, they can directly map a complete four-dimensional dispersion spectrum with the assistance of a wide-angle detector array and rotation of single crystal sample. Therefore, in terms of measuring a full four-dimensional dispersion spectrum, pulsed inelastic spectrometers is more efficient than triple-axis spectrometers. However, a triple-axis spectrometer can more efficiently track the changes in lattice dynamics with temperature field, magnetic field, stress field, etc. along a specific straight line in the 4D momentum-energy space[44].

2) Energy resolution. Energy resolution is an important parameter for inelastic and quasi-elastic neutron scattering spectrometer. Taking the direct geometry spectrometer as an example, the energy resolution is related to the energy broadening of the moderator, the distance from the moderator to the sample, the distance from the sample to the detector, the frequency of the monochromatic chopper, the energy of the incident neutron, and the energy transfer value, etc.[65] The energy resolution of a neutron scattering spectrometer is generally described by the ratio between the energy width at the elastic scattering position and the incident neutron energy, that is, $\Delta E/E_i$. The energy resolution of most direct geometry spectrometers is between 3% and 10%, and the resolution of a few spectrometers in the cold neutron energy range can reach ~1%. Taking the cold inelastic geometric spectrometer, AMATERAS at J-PARC in Japan, as an example, the resolution for an incident neutron energy of 3.134 meV can reach ~1.7%, and its absolute value is $\Delta E$ ~ 53 µeV. Regarding to inverse geometry spectrometers, they generally have a longer neutron flight distance, especially the sample-to-detector flight distance (Fig. 4(e)), which can provide a higher resolution. For example, the resolution of the BASIS spectrometer at SNS in the United States can be as high as ~0.1%, with an absolute value of about 3.5 µeV. However, the detector array of the inverse geometry spectrometer is concentrated around the sample, and the coverage area per unit solid angle is very small, so that its momentum resolution is much lower than that of the direct geometry spectrometer. According to the resolution characteristics of these two types of spectrometers, the direct geometry spectrometer is generally used to measure the phonon dispersion spectrum and the phonon density of states, while the inverse geometry spectrometer is generally used to study the behavior of ion diffusion and macromolecular vibration. It should be pointed out that the cold neutron direct geometry spectrometer is also often used to measure the dynamics of ion diffusion and macromolecular vibration because of its relatively higher energy resolution.

3) Quasi-elastic neutron scattering technique. As shown in Fig. 2(d), QENS is essentially an extreme case of INS. However, unlike INS, which is used to measure the collective dynamic behavior of lattice whose quantized energy deviates far from the elastic peak (except the acoustic phonon at the Brillouin zone center), QENS is mainly used to describe the non-quantized disordered motion such as diffusion and rotation in condensed matter[66,67]. The small energy transfer between neutrons and the disordered atoms moving in solid matter will cause Doppler broadening of the elastic peak. Therefore, QENS generally consists of two parts, an elastic scattering component described by the convolution of a $\delta$ function with the instrumental resolution, and a quasi-elastic part represented by the Lorentz peak adjacent to the elastic peak (Fig. 5(a)). In the study of energy materials, H, Li, Cu and Ag are the main elements that participate in disordered motion. These elements have large incoherent scattering cross-sections, and the related QENS signals are predominantly produced during incoherent scattering processes, which correspond to the self-motion of individual atoms or molecules rather than collective motion of the lattice.

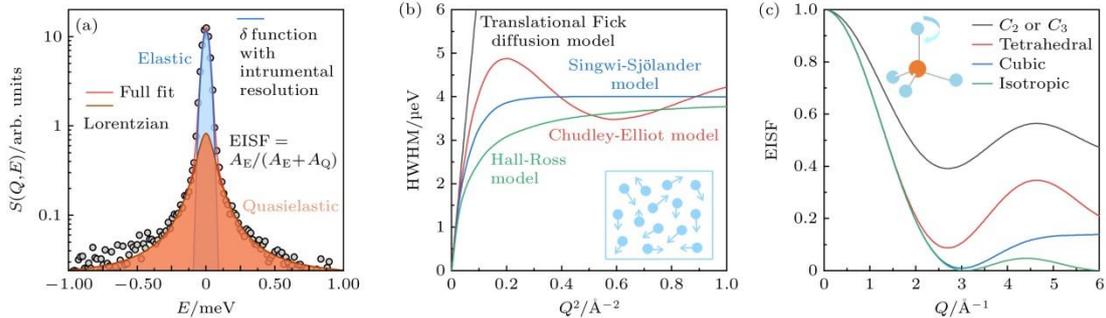

**Figure 5.** Data analysis methods for QENS data: (a) A general discomposing of a QENS spectrum, comprising an elastic part described by a $\delta$-function convoluted with instrumental resolution and a quasielastic part described with a Lorentzian profile, the ratio of the elastic area, $A_E$, to the total area, $A_E+A_Q$, defines the elastic incoherent structure factor (EISF)[66]; (b) the $Q$-dependent width of QENS signal can be used to study different ionic diffusion models[69]; (c) molecular rotation models under different geometric confinements and corresponding EISF profiles[70].

The elastic signal $S(Q)$ in the QENS spectrum carries information about the amplitude of the atomic vibration, i.e., mean square displacement $\langle u^2 \rangle$, and their specific correlation can be described by the formula $S(Q) \propto \exp(-Q^2 \langle u^2 \rangle/3)$[68]. The broadening of the quasi-elastic peak (half width at half maximum, HWHM) reflects the characteristic time of disordered motion. Based on the momentum dependence curve of HWHM, different modes of ion diffusion can be determined, such as the Fick model for long-distance diffusion, the Chudley Elliott model with a constant diffusion distance, and the Hall Ross model with a certain distribution of diffusion distance (Fig. 5 (b))[69] The elastic incoherent structure factor (EISF), which is composed of the ratio between elastic scattering intensity and total scattering intensity, is related to the geometric confinement of ion diffusion or molecular rotation, such as the rotation of ammonium tetrahedra along the $C_2/C_3$ axis or the cubic model of rotation between the eight vertices of a cube (Fig. 5(c))[70].

4) Other inelastic scattering techniques. In terms of four-dimensional dispersion spectroscopy, in addition to INS technology, Brillouin light scattering (BLS), inelastic X-ray scattering (IXS)[73,74] and four-dimensional electron energy-loss spectroscopy (4D-EELS)[75] can also be used. Regarding sample preparation, compared with the requirements of INS for large single crystal samples, the above three techniques are easier to meet the size requirements of single crystal samples, and some even do not require single crystal samples. In terms of lattice dynamics characterization, BLS can only measure acoustic phonons, and the energy range it can cover is very small (μeV level), while IXS and 4D-EELS can measure relatively complete dispersion spectra as INS. In terms of energy resolution, the resolution of IXS is generally on the order of meV, and the highest value reported so far can reach about 0.75 meV[74], but it is still not comparable to INS. The resolution of 4D-EELS is even worse, generally in the range of 14-16 meV, so this technique is mainly used for materials with relatively simple phonon spectrum structure and high energy[75]. However, among these inelastic scattering techniques, 4D-EELS has the highest spatial resolution and can achieve phonon spectrum characterization of thin films, surfaces, interfaces, local defects, etc. BLS, IXS and 4D-EELS techniques are similar to triple-axis neutron spectrometers in that they scan along a certain line in 4D space during dispersion spectrum measurement, which is different from direct geometry spectrometers with wide-angle detectors. The latter, due to the ability to simultaneously obtain lattice dynamics information of multiple complete Brillouin zones, can efficiently obtain full phonon dispersion when using single crystal samples and neutron-weighted (caused by different atomic nucleus scattering cross-sections) phonon density of states when using powder samples[21,76].

## 3. Ultralow lattice thermal conductivity in superionic thermoelectric materials.

Thermoelectric materials based on Seebeck effect and Peltier effect can convert heat energy into electricity through temperature gradient, which shows important application prospects in waste heat recovery, solid-state refrigeration, deep space exploration and other fields[77,78]. High-performance thermoelectric materials need as low lattice thermal conductivity as possible to ensure that the temperature gradient between the two ends of the device can be maintained during operation. The lattice thermal conductivity of materials can be described by the phonon free gas model $\kappa_{\text{lat}} = 1/3cv^2\tau = 1/3cvl$, where $c$, $v$, $\tau$, and $l$ represent the specific heat, phonon group velocity, phonon lifetime, and phonon mean free path, respectively[79]. Among these parameters, the phonon group velocity and lattice specific heat are related to atomic mass, chemical bond strength and unit cell size, and are intrinsic properties of materials, which are difficult to manipulate. In the development of thermoelectric materials, the common method to suppress the lattice thermal conductivity is to shorten the phonon lifetime or reduce its mean free path by introducing different scattering mechanisms, such as phonon anharmonicity, electron-phonon interaction, 0-dimensional point defect, 1-dimensional dislocation, 2-dimensional grain boundary scattering and multi-scale phonon scattering[9,79,80]. In recent years, some new methods to suppress the lattice thermal conductivity have been proposed, including lattice strain engineering, high entropy method and liquid-like phonon behavior[80–85]. Among them, superionic thermoelectric materials with liquid-like phonon behavior, involving complex crystal structures, ion migration, structural stability, phonon anharmonicity and disappearance of transverse phonon modes, are an important frontier research in the fields of condensed matter physics and energy materials.

In the microscopic lattice structure, this kind of superionic materials is composed of a stable rigid sublattice and a transportable liquid-like sublattice (Fig. 6(a))[85]. In such materials, the rigid sublattice can ensure good electrical properties (electronic crystal), while the liquid-like sublattice is considered to effectively suppress the lattice thermal conductivity (phonon liquid). In 2012, Liu et al. proposed the idea of using ionic liquid-like behavior to achieve low lattice thermal conductivity[23]. Up to now, many superionic thermoelectric materials satisfying the concept of "phonon liquid - electron crystal" have been discovered and studied, including $Cu_{2-\delta}$(S, Se, Te), $Ag_2$(S, Se, Te), (Ag, Cu)$CrSe_2$, $Cu_5FeS_4$ and argyrodite-based materials[85]. By optimizing the carrier concentration, controlling the lattice symmetry, and refining the grain size, the thermoelectric figure of merit of these materials can be increased to about 1.5 or even more than 2.0, which opens up a new field for the research of thermoelectric materials[84,85].

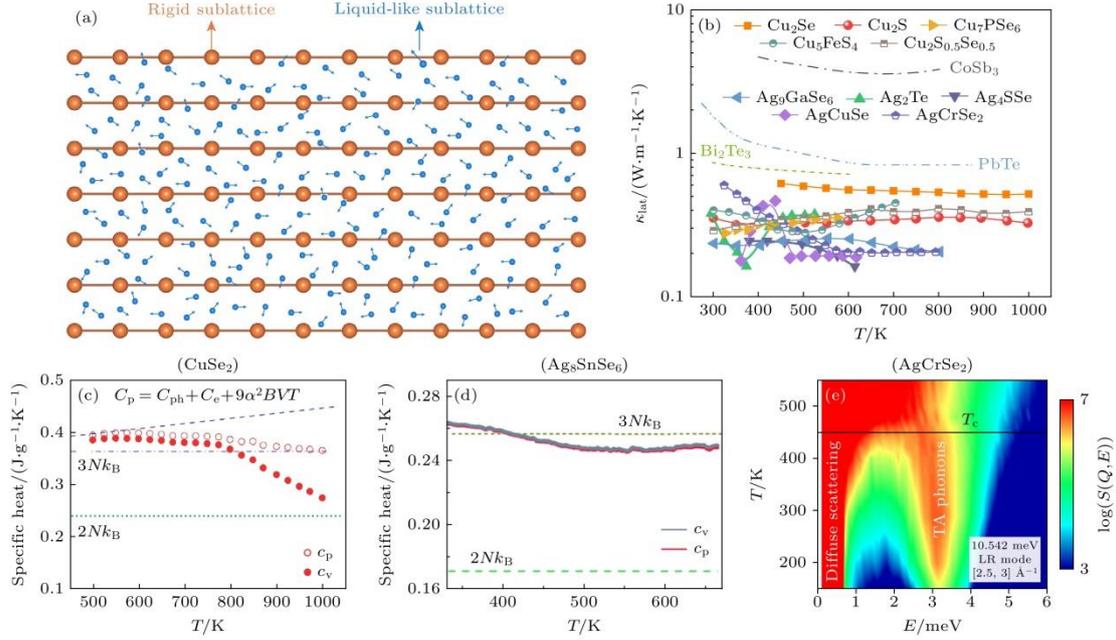

**Figure 6.** Superionic thermoelectric materials and ultra-low lattice thermal conductivity: (a) Schematic diagram of the crystal structure of superionic thermoelectric materials, comprising a rigid sublattice and a liquid-like sublattice[85]; (b) lattice thermal conductivity of several main superionic thermoelectric materials and a comparison with that for other typical thermoelectric materials[85]; (c)–(e) the attempts to demonstrate the validity of the liquid-phonon models for the ultralow lattice thermal conductivity through specific heat measurements and inelastic neutron scattering measurements on powder samples[23,24,52].

Compared with traditional thermoelectric materials, the biggest advantage of superionic thermoelectric materials is their extremely low lattice thermal conductivity. As shown in Fig. 6(b), the lattice thermal conductivity of this type of material is generally less than 1 W/(m·K), and can even be as low as 0.1-0.2 W/(m·K), which is 1-2 orders of magnitude lower than that of traditional thermoelectric materials[84,85]. However, there are a lot of controversy about the microscopic physical origin of the ultra-low lattice thermal conductivity in this type of material[10,18,19,23,29,52,86–89].

A popular view is that the vibration of liquid-like sublattice has no shear restoring force, the transverse phonon mode cannot be maintained, and the effective heat transport channel is reduced, resulting in a very low lattice thermal conductivity of the material[80]. Liu et al. found that the high temperature heat capacity at constant volume for $Cu_2Se$ and $Cu_8SnSe_6$ superionic materials is less than the Dulong-Petit limit, and used this phenomenon to try to prove the existence of liquid-like phonon behavior in such materials (Fig. 6(c),(d)). In 2018, Li et al observed the gradual disappearance of low-frequency phonons at ~3.2 meV around the superionic phase transition temperature in $CuCrSe_2$ powder samples by using inelastic neutron scattering technology, and attributed this behavior to the softening of transverse acoustic phonons (TA) and their integration into the diffuse scattering signal, thus trying to prove the important role of liquid-like phonons in ultra-low lattice thermal conductivity (Fig. 6(e))[52].

Nevertheless, the liquid-like phonon model was soon challenged. Whether the diffusion of ions affects the phonon vibration mode depends on the time scale comparison between these two dynamic behaviors. If the hopping time of atoms or ions between adjacent equilibrium positions in a liquid is shorter than the period of the lattice vibration, the phonon vibration mode will be destroyed. On the contrary, if the atom or ion stays longer than the vibration period, the phonon vibration and transmission will not be affected[90]. Voneshen et al. at Rutherford Appleton Laboratory in the United Kingdom first launched a challenge[29]. They used QENS technology to find that the diffusion jump time scale of Cu in $Cu_2Se$ compound is much smaller than the period of the general TA phonon, and only TA below 0.4 meV will be affected. In 2019, Xie et al. observed in real space using scanning transmission electron microscopy that Ag atoms in $AgCrSe_2$ have solid disorder behavior above the superionic phase transition temperature, rather than liquid behavior. In this case, many new explanations for the ultra-low lattice thermal conductivity of this type of materials have emerged, including low phonon group velocity[92], low acoustic phonon cut-off frequency[18], selective destruction of optical phonon[86], anharmonic phonon scattering[29] and lattice disorder scattering[91].

Regarding to the controversy over whether the liquid-like phonon model is the physical origin of the ultra-low lattice thermal conductivity of superionic compounds, the simplest and most direct approach is to measure the TA phonon before and after the superionic phase transition. To this end, Ren et al grew the argyrodite-based $Ag_8SnSe_6$ single crystal sample and used INS measurements to track the evolution of TA phonons along the [00$l$] direction in the (440) Brillouin zone. As shown in Fig. 7(a) and (b), the TA phonon is clearly measured at 300 K and 450 K, and the superionic phase transition does not effectively suppress the TA phonon vibration. This result directly disproves the liquid-like phonon model in which ion diffusion suppresses the thermal conductivity of TA phonon from the characterization of phonon spectrum. In order to reveal the main reason for the ultra-low lattice thermal conductivity in $Ag_8SnSe_6$, they measured the dynamic structure factor $S(Q,E)$ of $Ag_8SnSe_6$ powder sample at low temperatures by using the cold inelastic spectrometer AMATERAS ($E_i$ = 7.738 meV, $\Delta E$ ~ 0.223 meV) at J-PARC in Japan. At the extremely low temperature of 8 K, there are many dispersionless low-frequency optical phonons between 2 and 4 meV (Fig. 7(c)). However, these low-frequency phonons broaden rapidly with the increasing temperature, showing extremely large phonon anharmonic behavior (Fig. 7(d)-(f)). This rapid increase of phonon anharmonicity with temperature is consistent with the trend of rapid decrease of lattice thermal conductivity of $Ag_8SnSe_6$ sample from 20 to 50 K[31]. The microscopic lattice dynamics is highly consistent with the macroscopic transport properties, which indicates that the extremely shallow chemical energy surface and strong phonon anharmonicity in this material are the main physical mechanisms giving the sample ultra-low lattice thermal conductivity[31,93].

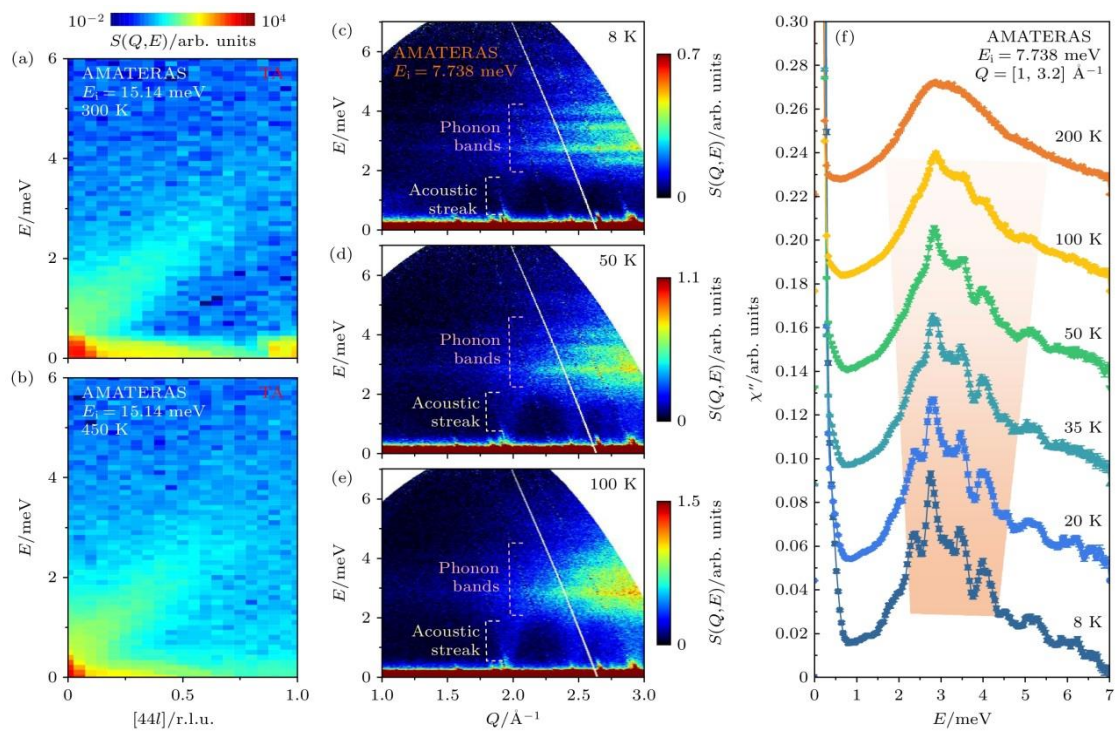

**Figure 7.** Lattice dynamics and ultra-low lattice thermal conductivity of $Ag_8SnSe_6$ argyrodite compounds[31]: (a), (b) TA phonons along the $[00l]$ direction in the (440) Brillouin zone of the $Ag_8SnSe_6$ single crystal samples measured on the cold neutron spectrometer AMATERAS at J-PARC at 300 K and 450 K, respectively; (c)–(e) dynamic structure factors $S(Q,E)$ of the $Ag_8SnSe_6$ powder samples at 8 K, 50 K and 100 K; (f) the peaks in the phonon density of states become broad quickly with increasing temperature.

## 4. Superionic phase transition and new mechanism of solid electrolyte

In battery devices, electrolyte is an important component, responsible for ion transport between the positive and negative electrodes, so that the battery form a closed current loop, and helps to maintain the electrical neutrality of the whole battery and prevent charge imbalance. The electrolyte of traditional batteries is liquid material, which has potential safety hazards such as leakage, volatilization and flammability. In order to overcome these shortcomings, the concept of solid-state battery has been proposed in recent years[94,95]. The essential difference between solid-state batteries and traditional batteries is that the liquid electrolyte is replaced by solid-state electrolyte. Solid-state batteries have the advantages of higher energy density, better safety, wider working range, longer cycle life and so on. They are an important research direction of ion batteries, and the development of solid-state electrolytes with high ionic conductivity is one of the core research contents[37].

The research of solid-state electrolytes faces many difficulties, including ion transport mechanism, electrochemical properties, mechanical properties and so on, among which the ion transport mechanism is the most basic scientific problem[36,37,96,97].

Ion diffusion channels and diffusion barriers are the main starting points for understanding the different ionic conductivities between different solid electrolytes, as shown in Fig. 8(a)[4,96,98]. However, a more fundamental physical understanding requires the consideration of the interaction between sublattices and of the lattice dynamic behaviors. In 2015, Wang et al. pointed out that the anion sublattice is closely related to the performance of transporting ions, and the body-centered cubic anion framework structure allows lithium ions to jump directly between adjacent tetrahedral sites, so it is the most ideal model to achieve high ionic conductivity[99]. In 2017, Kraft et al. studied the lattice stiffness of $Li_6PS_5X$ ($X$ = Cl, Br, I) through sound velocity measurement, and found that lattice softening can effectively reduce the diffusion barrier (Fig. 8(b))[100]. In 2018, Muy et al. further found through inelastic neutron scattering experiments that the lower the energy of low-frequency optical phonons, the lower the activation energy of ion diffusion (Fig. 8(c))[15]. More interestingly, in some materials, there is a phenomenon in which rotation of the polyanion sublattices drives the cation diffusion, which is called the "paddle-wheel model", as shown in Fig. 8(d)[101]. These previous studies show that the characterizations of crystal structure and lattice dynamics are important ways to study the ion transport mechanism of solid state electrolytes[102].

Although the important role of crystal structures and lattice dynamics in ion transport has gradually been recognized, previous studies were relatively macroscopic and lacked an understanding of the specific role of each atom. In-depth study of the unit cell will be more helpful for material design. Ren et al. used synchrotron XRD, neutron diffraction, INS and QENS techniques, combined with molecular dynamics simulation, to conduct in-depth and comprehensive study of the crystal structures, lattice dynamics, and ion diffusion of the typical $Ag_8SnSe_6$ argyrodites dynamics during superionic phase transition. The lattice dynamics data shows that phonons exhibit significant overdamping behavior above the superionic transition temperature, which is manifested as an additional scattering signal between the phonon peak and the zero-energy elastic peak in the dispersion spectrum (as indicated by the arrows in Fig. 9(a),(b)), while in the polycrystalline data, it is manifested as a rapid softening and broadening of the ~3 meV low-frequency phonon band, and its full width at half maximum is larger than the phonon center energy (see Fig. 9 (c)-(f)). This overdamping behavior of phonons means that some lattice vibration modes no longer require any energy to be activated, which coincides with the QENS broadening representing ion diffusion, indicating that the superionic phase transition is accompanied by a change in the correlation between the two sublattices, and this synergistic change can be reproduced by machine learning molecular dynamics simulations[31].

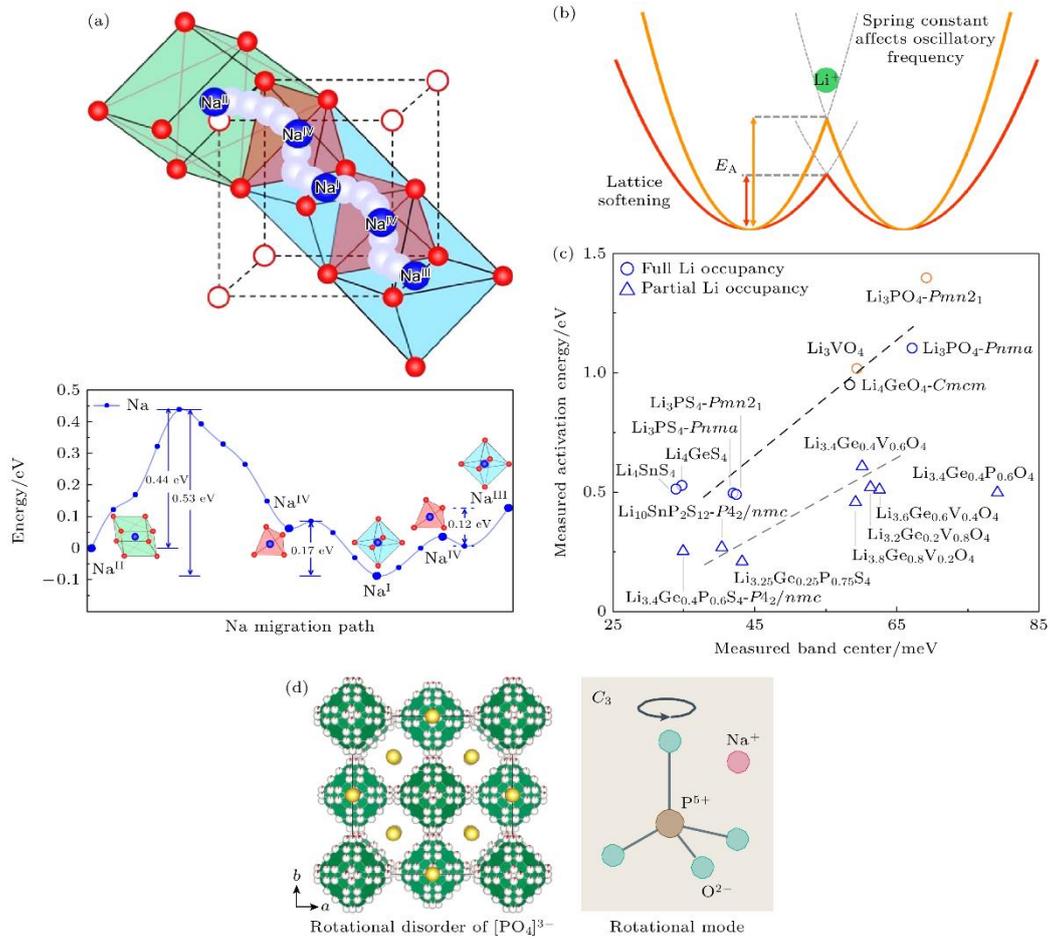

**Figure 8.** Ion diffusion and lattice dynamics of solid-state electrolytes: (a) Schematic diagram of ion diffusion channels and diffusion barriers in $Na_3Zr_2Si_2PO_{12}$[98]; (b) schematic diagram of the effect of lattice softening on hopping of ions[100]; (c) relationship between low-frequency phonon center energies measured by INS and ion diffusion activation energies in LISICON and olivine-type solid-state electrolytes[15]; (d) paddle wheel model of $Na^+$ ion transport in $\gamma$-$Na_3PO_4$ compounds, i.e., the rotation of $[PO_4]^{3-}$ polyanion tetrahedron drives the migration of $Na^+$ ions[101].

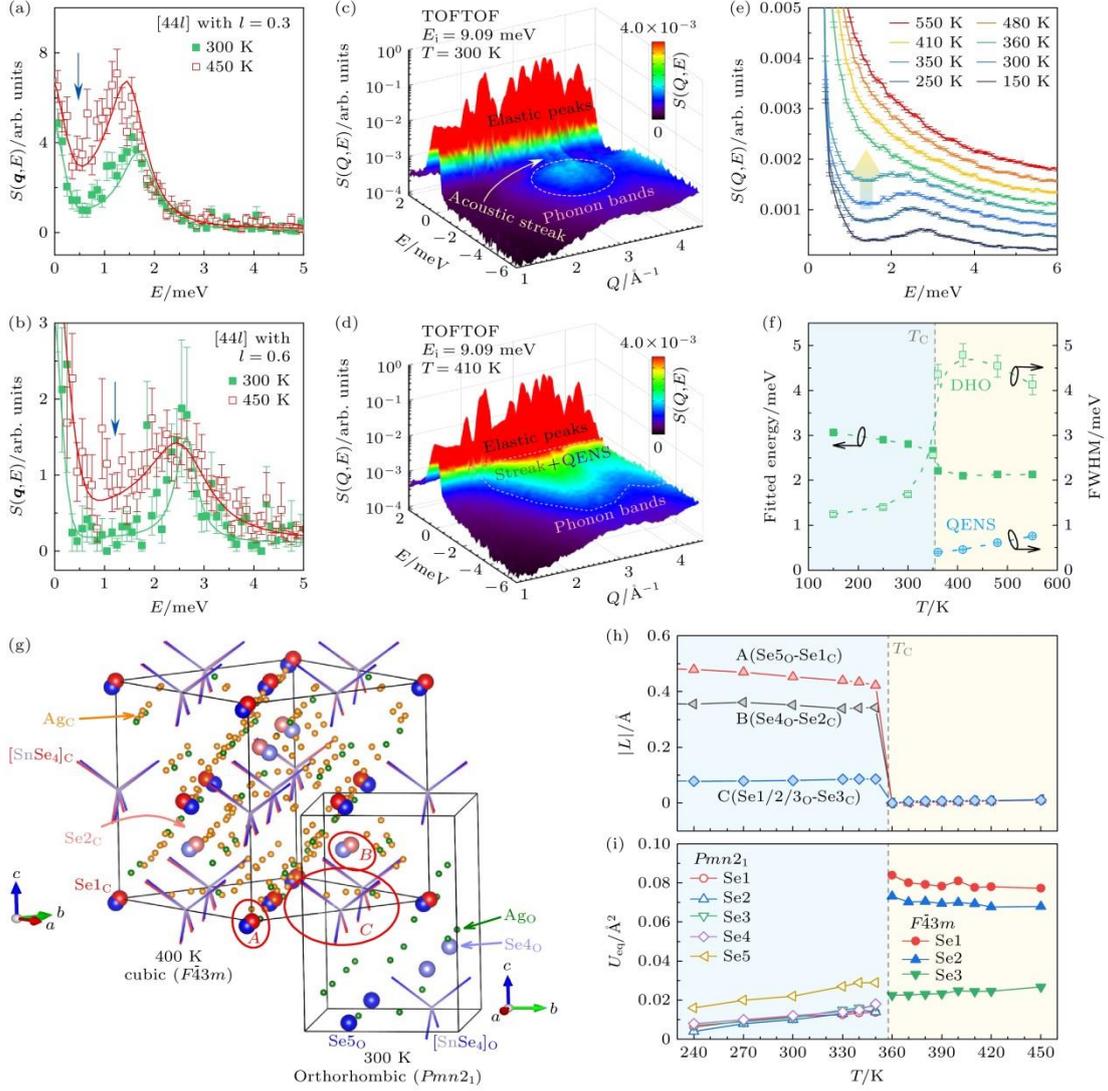

**Figure 9.** Lattice dynamics and coupling of sublattices during the superionic phase transition of $Ag_8SnSe_6$[31]: (a), (b) Experimental data of [4 4 0.3] and [4 4 0.6] TA phonon modes at 300 K and 450 K measured on the AMATERAS spectrometer at the J-PARC spallation neutron source in Japan, respectively; (c), (d) dynamical structure factors, $S(Q,E)$, of $Ag_8SnSe_6$ polycrystalline samples at 300 K and 410 K, respectively, measured on the TOFTOF direct-geometry spectrometer at the FRM II neutron source in Germany; (e) dynamical structure factor curves obtained by integrating $S(Q,E)$ over the momentum $Q$ range of [2.0, 2.4] Å$^{-1}$; (f) phonon energy, phonon width and QENS broadening obtained by fitting the curve in panel (e) with the damped harmonic oscillator (DHO) and Lorentz profiles; (g) comparison of the orthorhombic structure (O) and cubic structure (C) at 300 K and 400 K obtained on the D9 single-crystal neutron diffractometer at the ILL neutron source in France; (h), (i) the changes in the atomic displacement $|L|$ and isotropic atomic displacement parameters, $U_{eq}$, of Se atoms at positions *A*, *B*, and *C* during the superionic phase transition obtained by analyzing *in situ* XRD patterns, *A* corresponds to Se5 in the low-*T* phase and Se1 in the high-*T* phase, and *B* corresponds to Se4 in the low-*T* phase and Se2 in the high-*T* phase, respectively.

In order to further reveal the details of the correlation between the rigid sublattice and the superionic sublattice, the research group comprehensively compared the crystal structures before and after the phase transition. As shown in the Fig. 9(g),(h), the distribution of Ag ions in the superionic state becomes wider, and the Se atoms labeled by *A* and *B* undergo a large displacement, but the [SnSe$_4$] tetrahedron labeled by *C* does not change much during the phase transition (*A* corresponds to Se5 in the low temperature phase and Se1 in the high temperature phase, and *B* corresponds to Se4 in the low temperature phase and Se2 in the high temperature phase). This behavior indicates that the Se atoms at the *A* and *B* sites play a key role in the phase transition. To verify this inference, the changes of isotropic atomic displacement parameters $U_{eq}$ of Se atoms at different positions during the phase transition were compared in Fig. 9(i) by analyzing synchrotron XRD patterns. As expected by Fig. 9(g) and (h), the Se atoms at *A* and *B* sites have the largest changes in the $U_{eq}$ during the phase transition, and the machine learning molecular dynamics simulation results are consistent with the experimental results. These results show that the Se atoms at *A* and *B* play a central role in the sublattice interaction, which is an important modification site for regulating the ionic conductivity of this type of solid-state electrolyte materials, and it is also an important reference factor for improving the chemical stability when they are used as thermoelectric materials.

## 5. Plastic-crystal phase transition and configurational entropy for barocaloric refrigeration

Refrigeration is widely used in daily life and industrial production, accounting for more than 25% of the total energy consumption of the whole society[103]. At present, the widely used refrigeration equipment is mainly based on gas compression refrigeration technology, which has low energy efficiency and produces environmental problems such as greenhouse gases and ozone depletion[104,105]. Thermal caloric effects based on solid-state phase transitions, including magnetocaloric, electrocaloric, elastocaloric and barocaloric effects, are considered to be a new type of refrigeration technology that can improve energy efficiency and reduce environmental impact[106–113]. These thermal caloric effects have a common physical basis, namely the total enthalpy change or total entropy change during the corresponding phase transition, including lattice entropy, magnetic entropy, dipolar entropy, etc. Therefore, the characterization of crystal structure and lattice dynamics is an important way to study the mechanism of caloric materials[16,42,114,115].

The study of barocaloric materials is a typical example. Barocaloric effect is a physical phenomenon in which phase transition is driven by external pressure to realize storage and release of heat energy. Excellent barocaloric materials need as large phase transition entropy as possible, small hysteresis loss, and small driving pressure. In the pursuit of large phase transition entropy, plastic-crystal materials are a class of excellent candidate materials[116–120]. Plastic-crystal materials are a type of molecular crystals combined by weak long-range forces, and have the characteristic of long-range order and short-range disorder[16,101,121]. Unlike the translational symmetry breaking of the disordered ion distribution in superionic materials, the short-range disorder in plastic-crystal materials is rotational symmetry breaking, that is, the center of mass of the molecular units that constitute the plastic crystal materials are long-range ordered, but the spatial geometric orientation is disordered. This type of molecular orientational disorder can produce large configurational entropy during the phase transition, which is an ideal candidate for barocaloric refrigeration.

The conventional methods for studying the plastic-crystal phase transition and barocaloric effect include differential scanning calorimetry, XRD, *in-situ* Raman spectroscopy, et al.[119122123] In comparison, neutron scattering techniques can provide more comprehensive information[42120]. Li et al. found a giant barocaloric effect in a series of plastic crystal materials, which is one order of magnitude higher than that of traditional materials, and used neutron scattering technology to study the lattice dynamic behavior of neopentylglycol (NPG) plastic-crystal compounds[42]. The broadening of the phonon peak with temperature reveals that the material has a large lattice anharmonicity, which can endow the material with a large (ompressibility. QENS data and EISF analysis indicate that NPG molecules as a whole exhibit isotropic orientation rotation in the plastic-crystal state, providing the material with a large configurational entropy. At the same time, the inhibition of molecular orientation rotation by external pressure was observed by in situ pressurized QENS experiments.

In addition to the large phase transition entropy, the driving pressure of the phase transition that is as small as possible is also an important property for high-performance barocaloric refrigeration. Ren et al. systematically studied the phase transition and lattice dynamics of $NH_4I$ giant barocaloric compound by means of INS and QENS techniques. As shown in Fig. 10(a) and (b), the dynamic structure factor $S(Q,E)$ of the compound shows a significant quasi-elastic broadening with increasing temperature, indicating that the $[NH_4]^+$ tetrahedron has rotational behavior in the intermediate and high temperature phases, but the different degree of broadening indicates that the rotational behavior in the two phases is not the same. At the same time, the optical phonons undergo obvious anharmonic softening and broadening, especially near the intermediate-high temperature phase transition, the phonon peak broadens to the point that it can no longer be identified (Fig. 10(c)). This synergistic change between lattice vibration and molecular orientation rotation indicates that there is a significant coupling between the molecule and the lattice framework. ESIF and symmetry analysis further indicate that the $[NH_4]^+$ tetrahedron has two-fold spatial orientational degrees of freedom in the intermediate temperature phase and six-fold degrees of freedom in the high temperature phase. The schematic crystal structures are given in Fig. 10(d).

Here, the spatial orientation degree of freedom refers to the number of energetically degenerate spatial orientations of [NH$_4$]$^+$ tetrahedrons with relatively lower symmetry in the I$^-$ framework sublattice with relatively higher symmetry. This number depends on the N-H···I hydrogen bonds and the relative symmetries. In NH$_4$I, the change of rotational geometry also causes a change in the number of hydrogen bonds between [NH$_4$]$^+$ and I$^-$ (1↔4), which further strengthens the coupling between lattice vibration and molecular orientation rotation. This extremely strong coupling between sublattices makes the plastic-crystal phase transition in NH$_4$I very sensitive to pressure. The pressure dependence of the transition temperature is as high as $|dT_t/dP|$ ~0.79 K MPa$^{-1}$, endowing NH$_4$I with a very small saturation reversible driving pressure (~ 40 MPa).

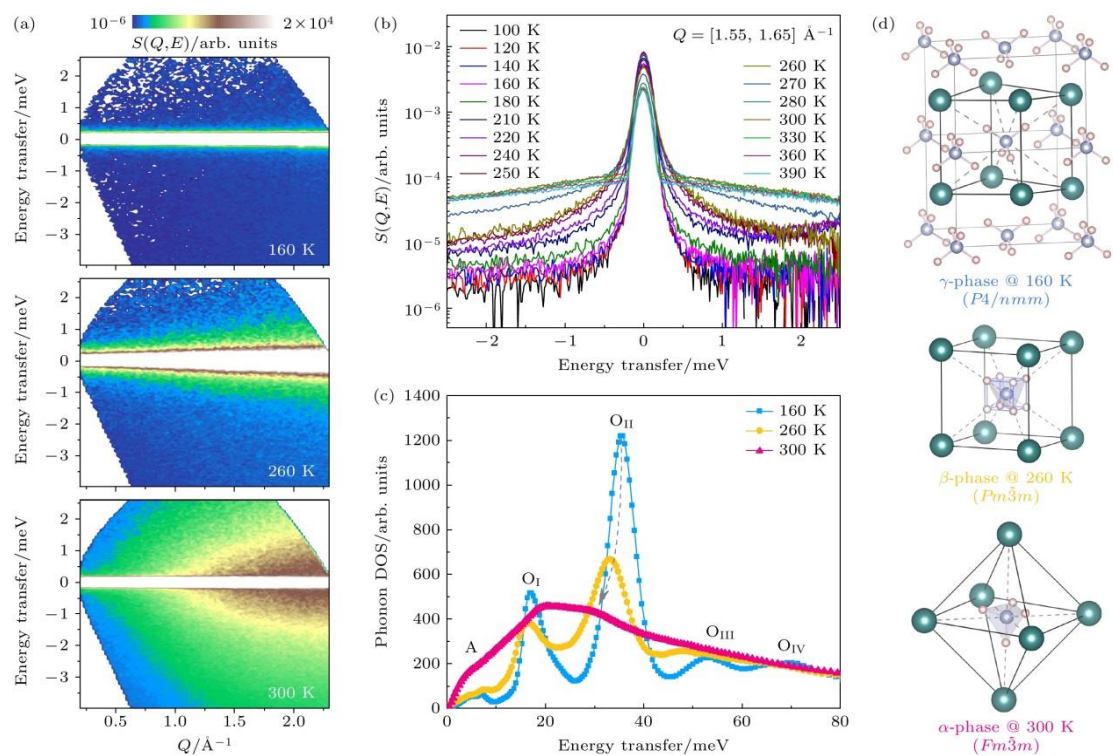

**Figure 10.** Plastic-crystal phase transition and lattice dynamics in NH$_4$I barocaloric material[68]: (a) Dynamic structure factors, $S(Q,E)$, in three different phases measured by INS; (b) quasielastic broadening in different phases obtained by integrating $S(Q,E)$ in (a) over a certain $Q$ range; (c) softening and broadening processes of the phonon DOSs with temperature; (d) crystal structures in the three different phases and possible orientations of the [NH$_4$]$^-$ tetrahedron.

## 6. The coupling between lattice and charge and spin

The above studies on lattice dynamics of superionic thermoelectric materials, solid electrolytes, and plastic crystal barocaloric materials all focus on the interaction between sublattices, but the coupling effect in energy materials goes far beyond this. In halogen perovskite optoelectronic materials, the lattice generally has very large anharmonicity, which is manifested as low thermal conductivity, large thermal expansion coefficient, and large atomic displacement parameter[124–126]. Lanigan-Atkins et al. used INS to observe that $CsPbBr_3$ single crystal form continuous diffuse rod-like network bands in three-dimensional reciprocal space (Fig. 11(a))[40]. These diffuse signals are caused by the damping behavior of low-energy phonons dominated by [$PbBr_6$] sublattice (Fig. 11(b)). Further studies have found that this giant anharmonic vibration of the lattice can modulate the electron-phonon coupling, thereby directly affecting the band-edge electronic states near the gap and making the hot carriers in this type of materials have a longer lifetime (Fig. 11(c))[40,127].

The electron-phonon interaction also plays an important role in thermoelectric materials. Ren et al. found in ZrNiSn-based half-Heusler alloy that the increase in carrier concentration can effectively shield the polarization electric field caused by the polar optical phonon vibration and hence increase carrier mobility[128]. This polarization electric field can lift the frequency of longitudinal optical phonon, causing longitudinal-transverse optical phonon splitting, and scatter carriers and reduce the mobility. Such shielding effect of carriers on the polarization electric field can also cause a significant softening of optical phonons in NbFeSb-based alloys, resulting in a significant reduction of lattice thermal conductivity[34].

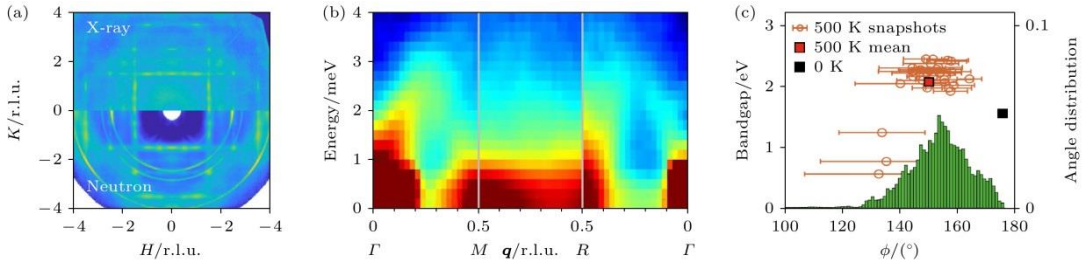

**Figure 11.** Relationship between lattice vibration and energy band in $CsPbBr_3$[40]: (a) Diffuse scattering signal in the (*H*, *K*, *L* = 0.5) plane in the high-temperature cubic phase at 433 K, the upper half of the data is obtained from XRD, and the lower half is from neutron data; (b) *S*(*Q*,*E*) along the *Γ-M-R-Γ* direction in the high-temperature phase at 419 K, where the phonons have overdamping behavior along the M-R direction at the Brillouin zone boundary; (c) Pb-Br-Pb angle distribution obtained by molecular dynamics simulation and its relationship with the energy band gap.

In the study of magnetocaloric refrigeration materials, first-order magneto-structural phase transition has attracted much attention because application of an external magnetic field can induce both magnetic entropy change and lattice entropy change simultaneously.

MnCoGe-based alloy compounds have relatively stable magnetic ordering temperature and easily controllable martensitic transformation temperature. When the magnetic phase transition coincides with the structural phase transition, spin-lattice coupling will be formed, thus realizing an external field-driven first-order magneto-structural transition[133–137]. Understanding and controlling the first-order magneto-structural transition is the focus of research on MnCoGe-based magnetocaloric materials. Ren et al. revealed that the lattice response of MnCoGe compound to magnetic field is caused by the elongation of Mn magnetic sublattice along the *c*-axis ($d_1$) and the shortening along the *a*-axis ($d_2$), which leads to the change in the framework of Co-Ge sublattice and then the entire crystal structure (Fig. 12(a)). By introducing additional valence electrons via Ni doping at the Co site, the martensitic transformation temperature ($T_M$) of MnCo$_{1-x}$Ni$_x$Ge alloys can be effectively modulated, decreasing from ~370 K at x = 0.18 to ~290 K, and then starting to increase at x = 0.55. The incorporation of Ni also significantly affects the magnetic structure of the low-temperature orthorhombic phase: the ferromagnetic alignment along the *c*-axis in low-doping regions first undergoes spin reorientation to a *b*-axis-oriented ferromagnetic configuration, followed by a transition to an incommensurate helical antiferromagnetic structure at x = 0.55 (Fig. 12(b)). The synergistic evolution of crystal and magnetic structures at x = 0.55 can be well explained by the transfer and redistribution of valence electrons between the Mn and Co-Ge sublattices, providing new insights for regulating spin-lattice coupling and magneto-structural transitions in such compounds[139,140].

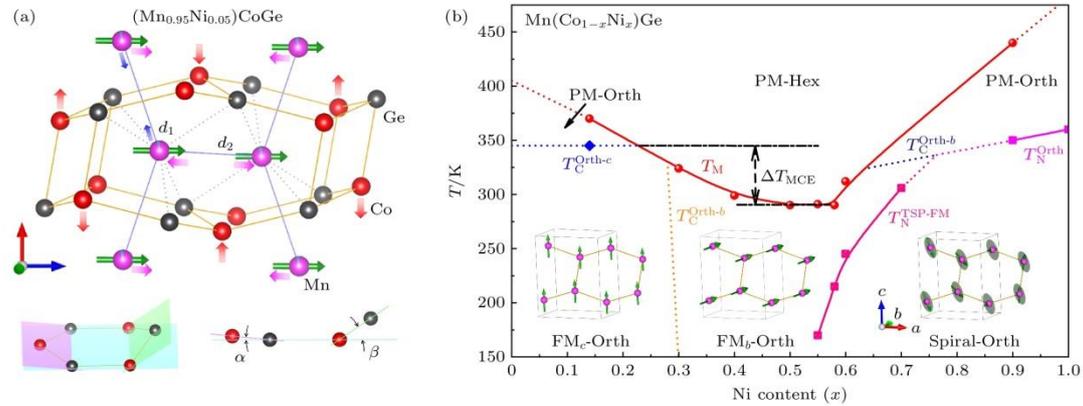

**Figure 12.** Magneto-structural transitions in MnCoGe-based magnetocaloric materials: (a) Response of the Mn magnetic sublattice and Co-Ge skeleton sublattice in the Mn$_{0.95}$Ni$_{0.05}$CoGe compound to an applied magnetic field obtained by *in situ* neutron diffraction experiments[138]; (b) magnetic structural phase diagram of the Mn(Co$_{1-x}$Ni$_x$)Ge compound with Ni content or valence electron number (Co: 3d$^7$4s$^2$, Ni: 3d$^8$4s$^2$) and temperature[139].

## 7. Summary and Prospect

Through the above five classic cases, it is not difficult to find that the microscopic crystal structure and lattice dynamics are the basis for understanding the complex physical mechanism of energy materials, and neutron scattering technology is an important spectroscopic means to characterize these information because of its high momentum and energy resolution. In superionic thermoelectric materials, the suppression of lattice thermal conductivity by giant phonon anharmonic scattering is more important than that by superionic phase transition and liquid-like phonon model. The weakening of the coupling between the cation and the rigid framework sublattice and the consequent overdamped vibration of the cation sublattice constitute an important dynamic behavior of ion diffusion in the ion transport process of the argyrodite-based solid state electrolyte. In plastic-crystal barocaloric materials, the enhancement of lattice anharmonicity with temperature can decouple the hydrogen bonds between ammonium molecules and the lattice framework, thus inducing a plastic-crystal phase transition and bringing about a large configurational entropy change. In addition, the interaction between lattice and charge makes it possible to improve the performance of photovoltaic and thermoelectric materials, and creating coupling between lattice and spin is an important strategy to design magnetocaloric materials. It can be seen that lattice dynamics does not exist independently in these energy conversion and storage materials, rather its complex role in macroscopic physical properties is always realized through the anharmonic evolution of phonons, combined with its unique multi-sublattice structure and the correlation with other degrees of freedoms, including sublattice, charge, spin, and so on.

Although abundant achievements have been made in the neutron scattering study of complex lattice dynamics of energy materials, there are still many scientific problems that need to be further understood. For example, in superionic thermoelectric materials, it has gradually become a consensus that anharmonic phonon scattering is the main physical origin of ultra-low lattice thermal conductivity[29,31,86], but how anharmonic scattering affects phonon transport in these materials has not been fully studied. In addition, ultra-low lattice thermal conductivity materials usually show unconventional temperature dependence[141–143]. At present, a popular explanation is that there are two channels for lattice thermal transport in such materials, one is the propagation of conventional phonon wave packets, and the other is the of random diffusion between different phonon branches[144–147]. Theoretically, the two-channel model corresponds to the of diagonal and off-diagonal matrix elements of phonons, respectively[146]. However, there is currently no clear method to verify or study this two-channel model from the perspective of atomic-level dynamic characterization using neutron spectroscopy. In the aspect of ion diffusion mechanism in solid electrolyte, it is theoretically analyzed that the diffusion barrier in the multi-ion concerted migration model is smaller than that in the single-ion migration model[96], but there are few experimental studies on multi-ion concerted diffusion. In the study of plastic-crystal barocaloric materials, phonon anharmonicity promotes the plastic-crystal phase transition, but also reduces the thermal conductivity, which hinders the application of materials in practical devices. Whether it is possible to decouple the plastic-crystal phase transition from thermal transport is an important scientific issue. In the aspect of magnetocaloric materials, the positive effect of first-order magneto-structural phase transition on phase transition entropy is based on the fact that the lattice entropy change and the magnetic entropy change absorb or release heat in the same direction during the phase transition process, but whether the two entropy changes contribute in the same direction in different materials and the proportion of their respective contributions are all issues that need to be considered in the material design process[115].

These new scientific problems in the study of the physical mechanism of energy materials listed above pose large challenges to the technologies of neutron scattering spectroscopy. Many materials have become extremely complex in crystallographic structure, such as superlarge unit cell, multiple doping, high entropy structure, organic-inorganic hybrid, partial order-partial disorder, grain nanocrystallization, micro-nano device interface, etc. These complex materials are often difficult to obtain large single crystal samples, and in some extreme cases, even if large single crystals can be obtained, micro-characterization is needed (such as grain nanocrystallization and micro-nano device applications), but it is extremely difficult to carry out micro-sample measurement or even micro-nano device measurement under the current limited neutron flux. Although IXS and 4D-EELS can be used to characterize the lattice dynamics of single crystal samples of millimeter scale or even smaller, the energy resolution of these techniques cannot meet the needs of complex lattice dynamics characterization in most cases. As a result, neutron scattering characterization of the phonon density of states has become one of the few options, but much of the useful information in the dispersion spectrum is lost in the "powder average". In the study of ion diffusion mechanism in solid electrolytes, how to distinguish coherent and incoherent scattering cross sections and quasi-elastic signals between different motion behaviors (such as ion diffusion and polyanion rotation) not only urgently needs to develop polarized inelastic neutron scattering techniques (polarized neutron techniques can distinguish coherent and incoherent scattering cross sections and nuclear and spin magnetic moment scattering), but also poses a high challenge to data processing methods and physical model analysis. Artificial intelligence and machine learning can automatically discover the rules hidden behind the data. It is known that they have achieved remarkable success in assisting the processing and analysis of electron microscopy data. The introduction of artificial intelligence and machine learning technology in the analysis of neutron scattering data of complex lattice dynamics may also provide some effective solutions to the above measurement and characterization difficulties.

As mentioned above, neutron scattering techniques, especially inelastic neutron scattering techniques, play an important role in the study of the physical mechanism of functional materials. At present, several triple-axis spectrometers have been built in two reactor neutron sources (CARR and CMRR) in China, and a high-energy direct geometry inelastic neutron scattering time-of-flight spectrometer has been built in the China Spallation Neutron Source (CSNS), which can meet many scientific research needs. Nevertheless, compared with the neutron sources in other countries, the current spectrometers are still not enough, especially in the cold neutron time-of-flight spectrometer, which is widely used in complex lattice dynamics spectroscopy because of its higher energy resolution. Fortunately, CSNS plans to build two new time-of-flight spectrometers in the cold energy range in the second phase of the "14th Five-Year Plan" project, which are expected to be completed around 2029. In terms of neutron flux, this CSNS Phase II project will upgrade the power from the current ~160 kW to ~500 kW, which is not up to the power level of J-PARC and SNS, but can greatly improve the experimental efficiency. In terms of inelastic neutron scattering user group, there are currently many professional research groups in the fields of superconductivity and quantum magnetism in China, but there are few in the fields of energy materials and phonons. Drawing on oversea experience, holding regular workshop on inelastic neutron scattering technology and scientific applications is an important way to quickly expand the user group and improve their skills. It is expected that after 10 to 20 years of development, the domestic inelastic neutron scattering technology and user groups will be significantly improved.